\documentclass[12pt]{article}

\usepackage{titling}
\usepackage[letterpaper, portrait, margin=1in]{geometry}
\usepackage{graphicx}
\usepackage{amsmath}
\usepackage{amssymb}
\usepackage{mathrsfs}
\usepackage[T1]{fontenc}
\usepackage{calligra}
\usepackage{array}
\usepackage{caption}
\usepackage{subcaption}
\usepackage{float}
\usepackage{bm}
\usepackage{color,soul}
\usepackage[toc,page,title]{appendix}
\usepackage[affil-it]{authblk}

\begin{document}

\title{Drude Conductivity of a Granular System}

\author{David T S Perkins and Robert A Smith}
\date{}
\affil{School of Physics and Astronomy, University of Birmingham, Edgbaston, Birmingham B15 2TT, United Kingdom}
\begin{titlingpage}
\maketitle
	
\begin{abstract}
We present a complete derivation of the granular analogue to Drude conductivity using diagrammatic methods. The convergence issues arising when
changing the order of momentum and frequency summation are more severe than in the homogeneous case. This is because there are now two momentum
sums rather than one, due to the intragrain momentum scrambling in tunnelling events. By careful analytic continuation of the frequency sum, and use
of integration by parts, we prove that the system is in the normal (non-superconducting) state, and derive the formula for the granular Drude conductivity
expected from Einstein's relation and Fermi's golden rule. We also show that naively performing the momentum sums first gives the correct result,
provided that we interpret a divergent frequency sum by analytic continuation using the Hurwitz zeta function.
\end{abstract}
	
\end{titlingpage}

\section{Introduction}\label{intro_sec}

The electrical conductivity of granular metals has been widely studied since the early 1960s, mainly in the vicinity of the metal-insulator transition,
using the ideas of variable-range hopping and percolation (see \cite{Abeles1975,Edwards_Rao_book} for reviews). However it is only relatively recently that
the diagrammatic techniques widely used in the analysis of homogeneous systems have been modified for use in granular metals. Varlamov et. al.
\cite{Varlamov1983} developed a diagrammatic description of a single S-N-S tunnel junction to consider the effect of superconducting fluctuations
on tunnelling current. Beloborodov et. al. \cite{Beloborodov1999,Beloborodov2000,Beloborodov2007} extended this approach to the model of a granular 
metal as a lattice of grains connected by weak links. Biagini et. al. \cite{Biagini2005_WL} used a similar approach to find the weak localisation correction to conductivity in a granular metal. These works all quote the granular analogue of the Drude conductivity in $d$ dimensions to be $\sigma_0^T=2e^2 N(0)\Gamma a^2$, where $N(0)$ is the single spin electronic density of states per unit volume at the Fermi surface, $\Gamma=2\pi N(0)t^2a^d$ is the electron
tunnelling rate, $t$ is a typical tunnelling matrix element, and $a$ is the average size of a grain. We have not been able to find an explicit diagrammatic
calculation of $\sigma_0^T$, although a path integral derivation is provided in the paper of Efetov and Tschersich \cite{Efetov2003}.

In this paper we provide a complete diagrammatic derivation of the granular Drude conductivity, $\sigma_0^T$, by analogy to the standard derivation
of Drude conductivity in a homogeneous system. Although there is no doubt of the validity of the formula for $\sigma_0^T$, reproducing this result
is an important test of the granular diagrammatic method. Moreover, there are convergence issues associated with changing orders of frequency and
momentum integrals in the homogeneous case, which need to be treated carefully to cancel the diamagnetic response. This leads us to expect similar
problems to arise in the granular system.

For simplicity of computation of electrical conductivity, it is preferable (when allowed) to perform the momentum integrals first, followed by the sum 
over Matsubara frequencies. In the Drude case it is necessary to first perform the Matsubara frequency sums by contour integration, and only then the
momentum integrals. We perform both ``naive'' calculations in which the orders of frequency and momentum integration are swapped without concern,
and more rigorous calculations in which convergence issues are treated carefully.

\section{Drude Conductivity in a Homogeneous Metal} \label{Homogeneous_sec}
	
This section will provide an overview of the calculation of Drude conductivity in a homogeneous metal using diagrammatic techniques, closely following the
presentation of refs \cite{Rickayzen,Altland_Simons}. Here we highlight the important details of constructing a rigorous diagrammatic theory for electrical
conductivity, and emphasise where issues of convergence occur in the simplest (Drude) calculation. This allows us to identify where similar difficulties are
likely to arise in the granular analogue of this calculation.
	
In general we can associate the current response of a system, $\mathbf{J}$, to the applied electric field, $\mathbf{E}$, via the conductivity tensor, 
$\sigma_{\alpha\beta}$, using
\begin{equation}
	J_{\alpha}(\mathbf{r},t) = \sum_{\beta}\int d^dr' \sigma_{\alpha\beta}(\mathbf{r-r'},t)E_{\beta}(\mathbf{r'},t).
	\label{current_conductivity_relation}
\end{equation}
Working in the Coulomb gauge, we consider a system in an electric field described by a vector potential, 
$\mathbf{E} = -\partial_{t}\mathbf{A}(\mathbf{r},t)$. Performing a temporal Fourier transform gives
\begin{equation}
	J_{\alpha}(\mathbf{r},\omega) = -i\omega\sum_{\beta}\int d^dr' \sigma_{\alpha\beta}(\mathbf{r-r'},\omega)A_{\beta}(\mathbf{r'},\omega),
	\label{current_conductivity_frequency_relation}
\end{equation}
so that the electrical conductivity is related to the linear response of the current to the vector potential.
	
The Hamiltonian for a disordered homogeneous metal in $d$ dimensions is
\begin{equation}
	H = \sum_{\sigma}\int d^{d}r\,\psi^{\dagger}_{\sigma}(\mathbf{r})\left[\frac{(-i\nabla-e\mathbf{A})^{2}}{2m} 
            + U(\mathbf{r})\right]\psi_{\sigma}(\mathbf{r}),
	\label{homogeneous_Hamiltonian}
\end{equation}
where $U(\mathbf{r})$ is the impurity potential, $m$ is the electron mass, and $\psi_{_{\sigma}}(\mathbf{r})$ is the field operator for an electron with 
spin $\sigma$. To consider the linear response of the system to an applied electric field, we write
\begin{equation}
	H = H_{0} + H',\qquad \text{where} \qquad H' = -\int d^{d}r\,\mathbf{A}(\mathbf{r},t) \cdot \mathbf{j}(\mathbf{r},t),
	\label{first_order_expansion_homogeneous_Hamiltonian}
\end{equation}
$H_{0}$ is the Hamiltonian in the absence of a vector potential, and $\mathbf{j}(\mathbf{r},t)$ is the electric current density operator,
\begin{equation}
	\mathbf{j}(\mathbf{r},t) = \frac{e}{2m}\sum_{\sigma}
        \Big[\psi^{\dagger}_{\sigma}(\mathbf{r},t)\left(-i\nabla-e\mathbf{A}(\mathbf{r},t)\right)\psi_{\sigma}(\mathbf{r},t) 
        + \left\{(i\nabla-e\mathbf{A}(\mathbf{r},t))\psi^{\dagger}_{\sigma}(\mathbf{r},t)\right\}\psi_{\sigma}(\mathbf{r},t)\Big].
        \label{homogeneous_current_operator}
\end{equation}
The latter may be written as
\begin{equation}
	\mathbf{j}(\mathbf{r},t) = \mathbf{j}_{0}(\mathbf{r},t) - 
        \frac{e^{2}}{m}\sum_{\sigma}\psi^{\dagger}_{\sigma}(\mathbf{r},t)\psi_{\sigma}(\mathbf{r},t)\mathbf{A}(\mathbf{r},t),     
        \label{homogeneous_current_operator_convenient}
\end{equation}
where $\mathbf{j}_{0}$ is the current operator in the absence of $\mathbf{A}$.
	
The macroscopic current, $\mathbf{J}(\mathbf{r},t)$, is given by taking the thermal average of $\mathbf{j}(\mathbf{r},t)$, followed by the average 
over the ensemble of impurity distributions. Using Kubo's formula for linear response \cite{Rickayzen}, we find
\begin{equation}
	J_{\alpha}(\mathbf{r},t) = \langle j_{0\alpha}(\mathbf{r},t) \rangle_{0} - \frac{ne^{2}}{m} A_{\alpha}(\mathbf{r},t)
        - \sum_{\beta}\int_{-\infty}^{\infty} dt' \int d^{d}r'\,\mathcal{G}^{R}_{\alpha\beta}(\mathbf{r},t;\mathbf{r'},t') A_{\beta}(\mathbf{r},t'),
	\label{linear_response_homogeneous_current}
\end{equation}
where $n = \sum_{\sigma}\langle \psi_{\sigma}^{\dagger}(\mathbf{r})\psi_{\sigma}(\mathbf{r}) \rangle_{0}$ is the conduction electron
number density, and $\langle ... \rangle_{0}$ denotes averaging with respect to $H_{0}$ and the impurity distribution. The retarded current-current
Green's function in eq. \ref{linear_response_homogeneous_current} can be written as
\begin{equation}
        \mathcal{G}^{R}_{\alpha\beta}(\mathbf{r},t;\mathbf{r}',t') = -i\langle [j_{0\alpha}(\mathbf{r},t),j_{0\beta}(\mathbf{r}',t')] \rangle_{0}\Theta(t-t'),
	\label{homogeneous_retarded_GF}
\end{equation}
where $\Theta(x)$ is the Heaviside function.
We extended the lower limit of the $t'$ integral from zero to $-\infty$ to isolate the driven response of the system from any transient response due to
switching the applied field on. The first term in eq. \ref{linear_response_homogeneous_current} equals zero, as it is just the average current in the absence 
of an applied field.
	
This retarded Green's function depends only upon the time difference $t-t'$, and we may therefore perform a temporal Fourier transform on it. The impurity
averaging leads to the system becoming translationally invariant, so we may also perform a spatial Fourier transform on 
eq. \ref{linear_response_homogeneous_current} to obtain
\begin{subequations}
\begin{equation}
	J_{\alpha}(\mathbf{q},\omega) = -\sum_{\beta}K_{\alpha\beta}(\mathbf{q},\omega)A_{\beta}(\mathbf{q},\omega),
	\label{current_linear_response_function_relation}
\end{equation}
\begin{equation}
	\hbox{where}\quad\quad
        K_{\alpha\beta}(\mathbf{q},\omega) = \frac{ne^{2}}{m}\delta_{\alpha\beta} + \mathcal{G}^{R}_{\alpha\beta}(\mathbf{q},\omega).
        \qquad\qquad\qquad
	\label{linear_response_function_homogeneous}
\end{equation}
	\label{Fourier_transformed_linear_response_homogeneous_current}
\end{subequations}
\par\noindent
The function $K_{\alpha\beta}(\mathbf{q},\omega)$ is known as the electromagnetic response function. From this we can derive the conductivity tensor 
using $K_{\alpha\beta}(\mathbf{q},\omega) = -i\omega\sigma_{\alpha\beta}(\mathbf{q},\omega)$.
	
The first term in $K_{\alpha\beta}(\mathbf{q},\omega)$ is known as the diamagnetic term, and is characteristic of superconducting behaviour. In a 
non-superconducting material, $\sigma_{\alpha\beta}(\mathbf{q},0)$ is finite, so that $K_{\alpha\beta}(\mathbf{q},0) = 0$. It follows that the 
diamagnetic term must be cancelled exactly by the zero-frequency current-current Green's function, $\mathcal{G}_{\alpha\beta}(\mathbf{q},0)$. 
In a superconducting material, $K_{\alpha\beta}(\mathbf{q},0)\ne 0$, and the diamagnetic term is not exactly cancelled by  
$\mathcal{G}_{\alpha\beta}(\mathbf{q},0)$.
	
We derive the retarded current-current Green's function by analytic continuation from its analogous temperature Green's function,
\begin{equation}
	\mathcal{G}_{\alpha\beta}(\mathbf{r}_{1},\tau;\mathbf{r}'_{1},\tau') = 
        -\langle T_{\tau}\left\{j_{0\alpha}(\mathbf{r}_{1},\tau),j_{0\beta}(\mathbf{r}'_{1},\tau')\right\} \rangle_{0}',
	\label{homogeneous_Matsubara_current_current_correlator}
\end{equation}
where $\tau$ is imaginary time, $T_{\tau}$ denotes imaginary time ordering, and $\langle ... \rangle_{0}'$ indicates thermal averaging over $H_{0}$ only. We will address the averaging over the impurity distribution shortly. Substituting $\mathbf{j}_{0}$ into $\mathcal{G}_{\alpha\beta}$ we obtain,
\begin{equation}
\begin{split}
	\mathcal{G}_{\alpha\beta}(\mathbf{r},\tau;\mathbf{r}',\tau') = \frac{e^{2}}{4m^{2}}\lim\limits_{\substack{\mathbf{r}'_{2} \rightarrow
        \mathbf{r}'_{1} \\ \mathbf{r}_{2} \rightarrow \mathbf{r}_{1}}} \bigg[&(\nabla_{2'}-\nabla_{1'})_{\beta} (\nabla_{2}-\nabla_{1})_{\alpha} \\
	&\times \sum_{\sigma,\sigma'} \langle T_{\tau}\big\{\psi_{\sigma}(\mathbf{r}_{2},\tau)
        \psi_{\sigma'}(\mathbf{r}'_{2},\tau')\psi^{\dagger}_{\sigma'}(\mathbf{r}'_{1},\tau')\psi^{\dagger}_{\sigma}(\mathbf{r}_{1},\tau)
        \big\rangle_{0}'\bigg].
	\label{current_current_correlator_limits}
\end{split}
\end{equation}
Applying Wick's theorem to the Green's function inside the limit we see there are two possible contractions,
\begin{equation}
	G_{0}(\mathbf{r}'_{2},\tau';\mathbf{r}'_{1},\tau')G_{0}(\mathbf{r}_{2},\tau;\mathbf{r}_{1},\tau) 
        - G_{0}(\mathbf{r}_{2},\tau;\mathbf{r}'_{1},\tau')G_{0}(\mathbf{r}'_{2},\tau';\mathbf{r}_{1},\tau)\delta_{\sigma\sigma'},
	\label{Wicks_theorem_homogeneous}
\end{equation}
where $G_{0}(\mathbf{r}_{1},\tau;\mathbf{r}'_{1},\tau') = -\langle T_\tau\big\{\psi_{\sigma}(\mathbf{r}_{1},\tau)\psi_{\sigma}^{\dagger} (\mathbf{r}'_{1},\tau')\big\} \rangle_{0}'$ is the single-electron Green's function in the absence of impurity averaging. The first term
of eq. \ref{Wicks_theorem_homogeneous} vanishes, as this corresponds to 
$\langle j_{0\alpha}(\mathbf{r}_{1},\tau) \rangle_{0}'\langle j_{0\beta}(\mathbf{r}'_{1},\tau') \rangle_{0}'$,
leaving just the second term of eq. \ref{Wicks_theorem_homogeneous}.
	
\begin{figure}[t]
	\centering
	\includegraphics[width=0.7\linewidth]{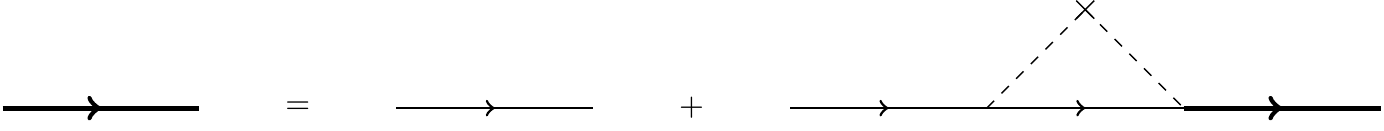}
	\caption{The leading order correction to the electron Green's function due to impurity averaging. The thin solid line is the free electron Green's
                     function; the thick solid line is the impurity dressed electron Green function; the dashed line denotes correlated impurity scattering events.}
	\label{impurity_GF_series}
\end{figure}

At this point we include the effects of impurity averaging, which we denote by $\langle ... \rangle$. This leads us to consider correlated impurity scattering
events within and between the electron Green's functions of eq. \ref{Wicks_theorem_homogeneous}. The leading order behaviour in a system with s-wave
scattering is found by neglecting correlated scattering events between the two Green's functions. We may therefore write
\begin{equation}
\begin{split}
	\langle G_{0}(\mathbf{r}_{2},\tau;\mathbf{r}'_{1},\tau')G_{0}(\mathbf{r}'_{2},\tau';\mathbf{r}_{1},\tau) \rangle &=
	\langle G_{0}(\mathbf{r}_{2},\tau;\mathbf{r}'_{1},\tau') \rangle \langle G_{0}(\mathbf{r}'_{2},\tau';\mathbf{r}_{1},\tau) \rangle \\
	&= G(\mathbf{r}_{2},\tau;\mathbf{r}'_{1},\tau')G(\mathbf{r}'_{2},\tau';\mathbf{r}_{1},\tau),
	\label{impurity_averaging}
\end{split}
\end{equation}
where $G(\mathbf{r}_{2},\tau;\mathbf{r}'_{1},\tau')$ represents the impurity averaged electron Green's function described by the diagrammatic series
in fig. \ref{impurity_GF_series}. We assume the impurity scattering events are correlated according to a white noise distribution,
\begin{equation}
\begin{split}
	\langle U(\mathbf{r},\tau) \rangle &= 0, \\
	\langle U(\mathbf{r},\tau)U(\mathbf{r}',\tau') \rangle &= \frac{1}{2\pi N(0)\tau_0}\delta^{(d)}(\mathbf{r}-\mathbf{r}')\delta(\tau-\tau'),
	\label{impurity_correlation}	
\end{split}
\end{equation}
where $1/\tau_0$ is the elastic scattering rate.
	
\begin{figure}[t]
	\centering
	\includegraphics[width=0.3\linewidth]{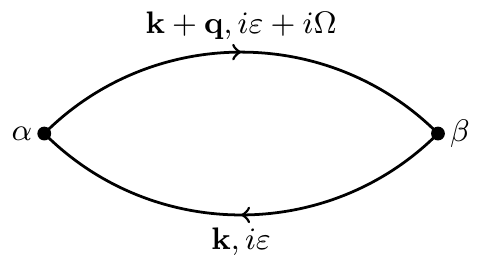}
	\caption{Diagrammatic representation of the homogeneous Drude linear response function in eq. \ref{Drude_diagram_GFs_full}.
                     The solid lines are the impurity averaged electron Green's functions; the solid dots are the current vertices.}
	\label{homogeneous_Drude_diagram}
\end{figure}

A consequence of the impurity averaging is to make the system translationally invariant, so that electron Green's functions depend only on the difference 
in their position variables, allowing us to make a spatial Fourier transform. Similarly, the fact that an electron Green's function depends only on the time
difference of its operators, allows us to make a temporal Fourier transform. The electron Green's function in momentum and frequency space is then \begin{equation}
	G(\mathbf{k},i\varepsilon) = \frac{1}{i\varepsilon - \xi_{\mathbf{k}} + \frac{i}{2\tau_0}\text{sgn}(\varepsilon)},
	\label{homogeneous_GF}
\end{equation}
where $\xi_{\mathbf{k}}=(k^2-k_F^2)/2m$, and $\varepsilon = (2l+1)\pi T$ ($l$ an integer) is a fermionic Matsubara frequency. The current-current Green's 
function in eq. \ref{current_current_correlator_limits} then becomes
\begin{equation}
	\mathcal{G}_{\alpha\beta}(\mathbf{q},i\Omega) = \frac{2e^{2}}{4m^{2}V} \sum_{\mathbf{k}}T\sum_{\varepsilon} 
        (2k_{\alpha} + q_{\alpha})(2k_{\beta} + q_{\beta}) G(\mathbf{k},i\varepsilon) G(\mathbf{k} + \mathbf{q},i\varepsilon + i\Omega),
	\label{Drude_diagram_GFs_full}
\end{equation}
where $\Omega = 2n\pi T$ ($n$ an integer) is a bosonic Matsubara frequency, and $V$ is the system's volume.  The current vertex in momentum space
naturally emerges from the Fourier transform as $e(2k_{\alpha}+q_{\alpha})/(2m)$. We represent eq. \ref{Drude_diagram_GFs_full} diagrammatically in fig. \ref{homogeneous_Drude_diagram}. The original retarded current-current Green's function is then obtained
from eq. \ref{Drude_diagram_GFs_full} by analytically continuing $i\Omega \rightarrow \omega + i\delta$, where $\delta$ is a positive infinitesimal.
	
In most cases of interest, $q\ll k_F$ and $\omega\ll E_F$, so that only electrons close to the Fermi surface are involved in transport. Consequently, the 
current vertices become $ek_{F\alpha}/m$, and $\xi_{\mathbf{k}+\mathbf{q}} \simeq \xi_{\mathbf{k}} + \mathbf{v}_{F} \cdot \mathbf{q}$ where 
$\mathbf{v}_{F}$ is the Fermi velocity. The momentum sum is approximated by
\begin{equation}
        \frac{1}{V}\sum_\mathbf{k}\approx N(0)\int d\hat{\Omega}_d \int_{-\infty}^\infty d\xi_k,
        \label{momentum_sum_approx}
\end{equation}
where $d\hat{\Omega}_d$ is the normalised $d$-dimensional angular element.

Here we note that issues of convergence arise. We cannot freely interchange the orders of summation over momentum and frequency, and so must proceed
carefully. We now approach this problem in two ways: first the naive approach where we arbitrarily swap the orders of summation; second a more
rigorous treatment of the swapping of summation order. From here onwards we shall consider a uniform electric field, so that we may set 
$\mathbf{q} = \mathbf{0}$.

\subsection{The Naive Approach} \label{Naive_homogeneous_sec}
Swapping the order of summation gives
\begin{equation}
\begin{split}
	\mathcal{G}_{\alpha\beta}(\mathbf{0},i\Omega) = &\frac{2e^{2}N(0)}{m^{2}} T\sum_{\varepsilon} \int d\hat{\Omega}_{d}\, 
        k_{F\alpha}\,k_{F\beta} \\
	&\times \int_{-\infty}^{+\infty} \frac{d\xi}{\left[\xi - i\varepsilon - \frac{i}{2\tau_0}\text{sgn}(\varepsilon)\right]
        \left[\xi - i\varepsilon - i\Omega - \frac{i}{2\tau_0}\text{sgn}(\varepsilon+\Omega)\right]}.
	\label{naive_response_order_swap}
\end{split}
\end{equation}
       For the $\xi$-integral to give a non-zero value, we need the two poles of the integrand to be in opposite half planes i.e. $\varepsilon$ and 
$\varepsilon+\Omega$ must have opposite sign. Assuming $\Omega > 0$ this leads to,
\begin{equation}
	-\Omega < \varepsilon < 0 \quad \Rightarrow \quad -\frac{\Omega}{2\pi T} - \frac{1}{2} < n < -\frac{1}{2}.
	\label{naive_frequency_range}
\end{equation}
Performing the $\xi$-integral using the method of residues gives
\begin{equation}
	\mathcal{G}_{\alpha\beta}(\mathbf{0},i\Omega) = \frac{2e^{2}N(0)}{m^{2}}\,T\sum_{\varepsilon}\Theta(-\epsilon(\epsilon+\Omega)) 
        \int d\hat{\Omega}_{d}\, k_{F\alpha}\,k_{F\beta}\,  \frac{2\pi\tau_0}{1+\Omega\tau_0}.
	\label{naive_post_momentum}
\end{equation}
The angular integral and frequency sum are straightforward, and yield factors of $k_{F}^{2}\delta_{\alpha\beta}/d$ and $\Omega/(2\pi T)$, 
respectively. Noting that the number density of electrons, $n = 4N(0)E_{F}/d$, we finally obtain
\begin{equation}
	\mathcal{G}_{\alpha\beta}(\mathbf{0},i\Omega) = \frac{ne^{2}\tau_0}{m}\frac{\Omega}{1+\Omega\tau_0}\delta_{\alpha\beta},
	\label{naive_bubble_result}
\end{equation}
so that the electromagnetic response function is
\begin{equation}
	K_{\alpha\beta}(\mathbf{0},i\Omega) = \frac{ne^{2}}{m} \delta_{\alpha\beta} 
        + \frac{ne^{2}\tau_0}{m}\frac{\Omega}{1+\Omega\tau_0}\delta_{\alpha\beta}.
	\label{naive_response_function}
\end{equation}
It is clear that $K_{\alpha\beta}(\mathbf{0},i\Omega)$ does not vanish in the limit $\Omega = 0$, and so it appears the system is superconducting rather
than normal. In evaluating $\mathcal{G}_{\alpha\beta}(\mathbf{0},i\Omega)$ we did not swap the order of summation with enough care, leading 
to the diamagnetic term in the response function not being cancelled.
	
If the electromagnetic response function was given by $\mathcal{G}_{\alpha\beta}(\mathbf{0},i\Omega)$ alone, we could use the relation 
$K_{\alpha\beta}(\mathbf{0},i\Omega) = \Omega\sigma_{\alpha\beta}(i\Omega)$ to obtain
\begin{equation}
	\sigma_{\alpha\beta}(\omega) = \frac{ne^{2}\tau_0}{m}\frac{1}{1-i\omega\tau_0}\delta_{\alpha\beta},
	\label{naive_AC_conductivity}
\end{equation}
after analytic continuation $i\Omega \rightarrow \omega + i\delta$. This is the well-known formula for the AC  Drude conductivity. However we still need
to understand how a rigorous treatment of the change of summation order leads to cancellation of the diamagnetic term.
	
\subsection{A Careful Treatment}

\begin{figure}[h]
	\centering
	\includegraphics[width=0.7\linewidth]{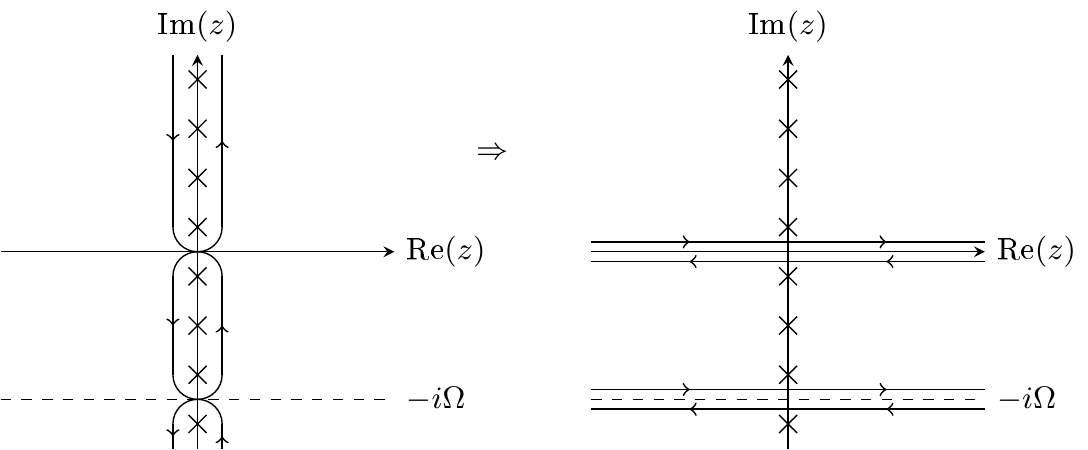}
	\caption{Analytic structure of the $\varepsilon$ sum in eq. \ref{Drude_diagram_GFs_full} before letting $i\Omega \rightarrow \omega + i\delta$. 
                     Branch cuts occur at $\text{Im}(z) = 0$ and $\text{Im}(z) = -\Omega$, due to the electron Green's functions $G(\mathbf{k},z)$
                     and $G(\mathbf{k+q},z+i\Omega)$, respectively.}
	\label{analytic_continuation_graph}
\end{figure}

We convert the Matsubara frequency sum in eq. \ref{Drude_diagram_GFs_full} into a contour integral in the complex frequency plane,
\begin{equation}
        T\sum_\varepsilon F(i\varepsilon)=-\frac{1}{2\pi i}\oint_C F(z)f(z)dz,
        \label{Matsubara_integral_relation}
\end{equation}
where $f(z)$ is the Fermi function, and $C$ is a contour enclosing the poles along the imaginary axis in the anticlockwise direction. We then deform the
contour as shown in fig.  \ref{analytic_continuation_graph} to pick up the branch cuts in the Green's function along $\text{Im}(z)=0$ and 
$\text{Im}(z)=-\Omega$, to give
\begin{equation}
\begin{split}
	\mathcal{G}_{\alpha\beta}(\mathbf{0},i\Omega) &= 
        \frac{e^{2}N(0)i}{\pi m^{2}} \int d\hat{\Omega}_{d}\,k_{F\alpha}\,k_{F\beta}\int_{-\infty}^{+\infty} d\xi \\
        &\qquad\times \Bigg\{ \int_{-\infty}^{+\infty} dz \Big[G^{R}(\mathbf{k},z) - G^{A}(\mathbf{k},z)\Big]G^{R}(\mathbf{k},z+i\Omega)f(z) \\
	&\qquad\qquad + \int_{-\infty-i\Omega}^{+\infty-i\Omega} dz \Big[G^{R}(\mathbf{k},z+i\Omega) - G^{A}(\mathbf{k},z+i\Omega)\Big]
        G^{A}(\mathbf{k},z)f(z) \Bigg\}.
	\label{Drude_response_pre_frequency_shift}
\end{split}
\end{equation}
We then shift $z \rightarrow z - i\Omega$ in the second frequency integral, noting that $f(z+i\Omega)=f(z)$, before analytically continuing 
$i\Omega \rightarrow \omega + i\delta$, to yield
\begin{equation}
\begin{split}
	\mathcal{G}_{\alpha\beta}^{R}(\mathbf{0},\omega) = &\frac{e^{2}N(0)i}{\pi m^{2}} \int d\hat{\Omega}_{d}\, k_{F\alpha}\,k_{F\beta} 
        \int_{-\infty}^{+\infty} d\xi \\
	&\times \int_{-\infty}^{+\infty} dz \Big[G^{R}(\mathbf{k},z) - G^{A}(\mathbf{k},z)\Big] 
        \Big[G^{R}(\mathbf{k},z+\omega) + G^{A}(\mathbf{k},z-\omega)\Big] f(z).
	\label{Drude_response_post_frequency_shift}
\end{split}
\end{equation}
	
We first consider the $G^{R}G^{R}$ term in eq. \ref{Drude_response_post_frequency_shift}, and initially set $\omega=0$, so that we have the factor
$G^R(k,z)^2$. The integral may then be written as
\begin{equation}
        \int_{-\infty}^{+\infty} d\xi \int_{-\infty}^{\infty} dz {1\over\left(z-\xi+{i\over 2\tau_0}\right)^2}f(z)=
        - \int_{-\infty}^{+\infty} d\xi \int_{-\infty}^{\infty}dz {d\over dz}\left[{1\over z-\xi+{i\over 2\tau_0}}\right] f(z)
\end{equation}
Performing the $z$-integral by parts generates a vanishing boundary term, and shifts the derivative onto the Fermi function. We can now swap the order 
of integration as $f'(z)$ falls off sufficiently rapidly at infinity to yield
\begin{equation}
        \int_{-\infty}^{+\infty}dz\,{df\over dz}\int_{-\infty}^{+\infty}{d\xi\over z-\xi+{i\over 2\tau_0}}=
        \int_{-\infty}^{+\infty}dz\,{df\over dz}(-\pi i)=\pi i.
\end{equation}
To consider $\omega\ne 0$, we expand $G^R(k,z+\omega)$ as a power series in $\omega$ giving
\begin{equation}
        G^R(k,z)G^R(k,z+\omega)=\sum_{n=0}^\infty (-1)^n {\omega^n\over\left(z-\xi+{i\over 2\tau_0}\right)^{n+2}}=
        {d\over dz}\sum_{n=0}^\infty {(-1)^{n+1}\over (n+1)} {\omega^n\over\left(z-\xi+{i\over 2\tau_0}\right)^{n+1}}
\end{equation}
Integrating by parts as before, and swapping the order of integration, we find that the $n\ne 0$ terms vanish upon integration over $\xi$, and we recover 
the same result $\pi i$ as in the case where $\omega=0$. An entirely equivalent procedure may be applied to the $G^AG^A$ term, so that
\begin{equation}
        \int_{-\infty}^{+\infty} d\xi \int_{-\infty}^{+\infty} dz\,G^A(k,z)\,G^A(k,z-\omega)\,f(z)=-\pi i.
\end{equation}
Thus the $G^{R}G^{R}$ and $G^{A}G^{A}$ contributions to eq. \ref{Drude_response_post_frequency_shift} may be written as
\begin{equation}
        \frac{e^{2}N(0)i}{m^{2}} \int d\hat{\Omega}_{d}k_{F\alpha}k_{F\beta}(2\pi i)=
        -\frac{2e^{2}N(0)k_{F}^{2}}{m^{2}d}\delta_{\alpha\beta} 
        = -\frac{ne^{2}}{m}\delta_{\alpha\beta},
        \label{retarded_advanced_contributions_homogeneous}
\end{equation}
which exactly cancels the diamagnetic term in $K_{\alpha\beta}(\mathbf{0},\omega)$.
	
The completely retarded and advanced pieces of eq. \ref{Drude_response_post_frequency_shift} therefore cancel the
diamagnetic term in $K_{\alpha\beta}(\mathbf{0},\omega)$ exactly. The $\mathcal{O}(\omega^{0})$ contribution of the 
$G^RG^A$ terms is zero since setting $\omega = 0$ causes these terms to cancel. Hence the linear response function,
$K_{\alpha\beta}(\mathbf{0},\omega)$, vanishes in the limit $\omega = 0$, as required for a normal metal.
	
If we now consider the $G^RG^A$ terms in eq. \ref{Drude_response_post_frequency_shift}, they may be combined by shifting 
$z \rightarrow z + \omega$ in the $G^{R}(\mathbf{k},z)G^{A}(\mathbf{k},z-\omega)$ piece to yield
\begin{equation}
	\int_{-\infty}^{+\infty}d\xi\int_{-\infty}^{+\infty} dz\,G^R(k,z+\omega)\,G^A(k,z)\Big[f(z+\omega)-f(z)\Big].
	\label{cross_terms_shift_homogeneous}
\end{equation}
The $[f(z+\omega) - f(z)]$ term falls off rapidly enough at infinity to change the order of integration. Performing the $\xi$ integral followed by the 
$z$ integral we find
\begin{equation}
	K_{\alpha\beta}(\mathbf{0},\omega)= 
        \frac{e^{2}N(0)i}{\pi m^{2}}\cdot\frac{-2\pi\omega\tau_0}{1-i\omega\tau_0} \int d\hat{\Omega}_{d}\,k_{F\alpha}\,k_{F\beta}
        =-{i\omega\over 1-i\omega\tau_0}{ne^2\tau_0\over m}\delta_{\alpha\beta},
	\label{homogeneous_response_simplified}
\end{equation}
where we used the result
\begin{equation}
        \int_{-\infty}^{+\infty} \Big[f(z)-f(z+\omega)\Big] dz=\omega.
\end{equation}
From the relation $K_{\alpha\beta}(\mathbf{q},\omega) = -i\omega\sigma_{\alpha\beta}(\omega)$, we obtain the AC conductivity tensor \cite{Rickayzen,Altland_Simons,AGD}
\begin{equation}
	\sigma_{\alpha\beta}(\omega) = \frac{ne^{2}\tau_0}{m} \frac{1}{1-i\omega\tau_0}\,\delta_{\alpha\beta},
        \label{homogeneous_AC_conductivity}
\end{equation}
and setting $\omega = 0$ we finally arrive at the well-known Drude conductivity formula
\begin{equation}
	\sigma_{0} = \frac{ne^{2}\tau_0}{m}.
	\label{Drude_conductivity}
\end{equation}
	
In this section we have provided a rigorous treatment of the electrical conductivity of a homogeneously disordered conductor using quantum field theory methods. Doing so has allowed us to carefully construct the diagrammatic rules associated to electrical conductivity calculations, as well as highlighting points where we must proceed with caution. We now consider how to model a granular system in a similar manner.
	
\section{Diagrammatic Theory for Granular Electrical Conductivity} \label{Diagrammatics_sec}
	
In this section we provide an analogous treatment for electrical conductivity in a granular metal to that of section \ref{Homogeneous_sec} for a
homogeneous metal. Once again we encounter issues of convergence when swapping the order of summation over frequency and momentum.
Naively swapping order and performing the momentum sum first leads to a more straightforward calculation, but to proceed rigorously we must
perform the frequency sum first. We show that the linear response function, $K_{\alpha\beta}(\omega)$ vanishes in the zero frequency
limit, and obtain the formula for the granular Drude conductivity.
	
We start from the Hamiltonian for a system of identical grains on a cubic lattice of side $a$ in the presence of a vector potential $\mathbf{A}$,
\begin{equation}
	H = \sum_{i}\sum_{\mathbf{k},\sigma}\xi_{i\mathbf{k}}c_{i\sigma\mathbf{k}}^{\dagger}c_{i\sigma\mathbf{k}}^{\null} + \sum_{i}\sum_{\substack{\mathbf{k},\mathbf{q}, \\ \sigma}} U_{i}(\mathbf{q})c_{i\sigma\mathbf{k}+\mathbf{q}}^{\dagger}c_{i\sigma\mathbf{k}}^{\null} +  \sum_{i,j}\sum_{\substack{\mathbf{k},\mathbf{p}, \\ \sigma}}t_{ij}^{\mathbf{k}\mathbf{p}}e^{ie\mathbf{A}\cdot \mathbf{R}_{ij}} c_{i\sigma\mathbf{k}}^{\dagger}c_{j\sigma\mathbf{p}}^{\null},
	\label{full_Hamiltonian}
\end{equation}
as considered by Beloborodov et. al. \cite{Beloborodov2007} and Biagini et. al. \cite{Biagini2005_WL}.
In the above, $\mathbf{R}_{ij} = \mathbf{R}_{i} - \mathbf{R}_{j}$, where $i$ and $j$ label the grains located at lattice sites $\mathbf{R}_{i}$ 
and $\mathbf{R}_{j}$ respectively, $\mathbf{k}$ and $\mathbf{p}$ are the electron momenta internal to a grain, $t_{ij}^{\mathbf{k}\mathbf{p}}$ are the tunnelling matrix elements associated to moving from the state $\mathbf{p}$ in grain $j$ to state $\mathbf{k}$ in grain $i$, $\xi_{i\mathbf{k}}$ is the internal energy of an electron in state $\mathbf{k}$ on the $i$\textsuperscript{th} grain, $U_{i}(\mathbf{q})$ is the potential due to the impurity distribution on the $i$\textsuperscript{th} grain, and $\sigma$ is the electron spin. We consider only nearest-neighbour
hopping, with random tunnelling elements which satisfy
\begin{equation}
\begin{split}
        \langle t_{ij}^{\mathbf{k}\mathbf{p}}\rangle&=0, \\
        \langle t_{ij}^{\mathbf{k}\mathbf{p}}t_{lm}^{\mathbf{k}'\mathbf{p}'}\rangle &=
        \begin{cases}
	        t^2(\delta_{im}\delta_{jl}+\delta_{il}\delta_{jm})\delta_{\mathbf{k}+\mathbf{k}'=\mathbf{p}+\mathbf{p}'}, \qquad i, \, j \, \text{ and } l, \, m \text{ nearest neighbours} \\
	        0, \qquad \text{otherwise.}
        \end{cases}
        \label{tunnelling_matrix_elements}
\end{split}
\end{equation}
Since we only have nearest-neighbour tunnelling, $j=i\pm\alpha$ where $\alpha=x, y, z$, etc. The tunnelling Hamiltonian may then be written as
\begin{equation}
        H_T=\sum_{i,\alpha}\sum_{\substack{\mathbf{k},\mathbf{p}, \\ \sigma}}
                \left( t_{i+\alpha,i}^{\mathbf{k}\mathbf{p}} e^{ieaA_{\alpha}} c_{i+\alpha\sigma\mathbf{k}}^{\dagger}c_{i\sigma\mathbf{p}}+
                t_{i-\alpha,i}^{\mathbf{k}\mathbf{p}}  e^{-ieaA_{\alpha}} c_{i-\alpha\sigma\mathbf{k}}^{\dagger}c_{i\sigma\mathbf{p}}\right),
        \label{tunnelling_Hamiltonian}
\end{equation}
where $A_{\alpha}$ is the component of $\mathbf{A}$ in the $\alpha$\textsuperscript{th} direction. The electrical current in the $\alpha$\textsuperscript{th} direction is
then
\begin{equation}
       j_{\alpha} = -\frac{1}{a^{d}\mathcal{N}}{\delta H\over\delta A_\alpha} = -\frac{ie}{a^{d-1}\mathcal{N}} \sum_{i}\sum_{\substack{\mathbf{k},\mathbf{p}, \\ \sigma}}\left(t_{i+\alpha,i}^{\mathbf{k}\mathbf{p}}
                          e^{ieaA_{\alpha}} c_{i+\alpha\sigma\mathbf{k}}^{\dagger}c_{i\sigma\mathbf{p}}-\hbox{H.c.}\right),
      \label{full_current_operator}
\end{equation}
where $\mathcal{N}$ is the number of grains in the system. An alternative derivation of this current operator $j_{\alpha}$, starting from the lattice version of the continuity equation, is given in appendix A.
	
To consider the linear response of the system, we expand $H_{T}$ and $j_{\alpha}$ to first order in $\mathbf{A}$ giving
\begin{subequations}
\begin{equation}
	H_{T} = H_{T}^{(0)} - a^{d}\mathcal{N}\sum_{\alpha}A_{\alpha}j_{0,\alpha},
	\label{linear_response_tunnelling_Hamiltonian}
\end{equation}
\begin{equation}
	j_{\alpha} = j_{0,\alpha} + \frac{e^{2}A_{\alpha}}{a^{d-2}\mathcal{N}}\sum_{i}\sum_{\substack{\mathbf{k},\mathbf{p}, \\ \sigma}}\left(
                           t_{i+\alpha,i}^{\mathbf{k}\mathbf{p}} c_{i+\alpha\sigma\mathbf{k}}^{\dagger}c_{i\sigma\mathbf{p}} + \hbox{H.c.}\right) 
                        = j_{0,\alpha} + \frac{e^{2}A_{\alpha}}{a^{d-2}\mathcal{{N}}}H_{T}^{(0)},
	\label{linear_response_current_operator}
\end{equation}
\label{linear_expansions}
\end{subequations}
where $H_{T}^{(0)}$ and $j_{0,\alpha}$ are the tunnelling Hamiltonian and current operator in the absence of a vector potential. 
From standard linear-response theory \cite{AGD}, we see that the macroscopic current is given by
\begin{equation}
	J_{\alpha}(t) = \langle j_{0,\alpha}(t) \rangle_{0} + \frac{e^{2}A_{\alpha}(t)}{a^{d-2}\mathcal{N}}\langle H_{T}^{(0)} \rangle_{0} 
                               - \sum_{\beta}\int_{-\infty}^{+\infty}dt' \mathcal{G}_{\alpha\beta}^{R}(t,t')A_{\beta}(t'),
	\label{macroscopic_current}
\end{equation}
where $\mathcal{G}_{\alpha\beta}^{R}(t,t')$ is the retarded current-current correlator,
\begin{equation}
	\mathcal{G}_{\alpha\beta}^{R}(t,t') = -i\langle \left[j_{0,\alpha}(t),j_{0,\beta}(t')\right] \rangle_{0}  \, a^{d}\mathcal{N} \, \Theta(t-t').
	\label{retarded_current_current_correlator}
\end{equation}
We have used $\langle ... \rangle_{0}$ to denote averaging with respect to $H_{T}^{(0)}$, as well as over the impurity distribution within the grains, 
and the tunnelling matrix element distribution.
	
Clearly $\langle j_{0,\alpha} \rangle_{0} = 0$, as the current in the absence of an applied field is zero. In a similar manner, 
$\langle H_{T}^{(0)} \rangle_{0} = 0$, as the average over $t_{i+\alpha,i}^{\mathbf{k}\mathbf{p}}$ gives zero. 
This leaves just the third term of eq. \ref{macroscopic_current} contributing to the macroscopic current. To analyse this term, we move to the 
Matsubara formulation of the current-current correlator,
\begin{equation}
	\mathcal{G}_{\alpha\beta}(\tau,\tau') = -\langle T_{\tau}\left\{j_{0,\alpha}(\tau)j_{0,\beta}(\tau')\right\} \rangle_{0}.
	\label{Matsubara_currect_current_correlator}
\end{equation}

Proceeding as before, we apply Wick's theorem to $\mathcal{G}_{\alpha\beta}(\tau,\tau')$, average over the distribution of tunnelling matrix elements, and expand the Green's functions as temporal Fourier series to find
\begin{equation}
	\mathcal{G}_{\alpha\alpha}(i\Omega) = \frac{2e^{2}t^{2}}{a^{d-2}\mathcal{N}} \sum_{i}\sum_{\mathbf{k},\mathbf{p}} T\sum_{\varepsilon} G_{i}(\mathbf{k},i\varepsilon+i\Omega) \left[G_{i+\alpha}(\mathbf{p},i\varepsilon) + G_{i-\alpha}(\mathbf{p},i\varepsilon)\right].
	\label{linear_response_function}
\end{equation}
To obtain eq. \ref{linear_response_function} we made use of
\begin{equation}
	-\langle T_{\tau}\{c_{i\sigma\mathbf{k}}^{\null}(\tau)c_{j\sigma'\mathbf{p}}^{\dagger}(\tau')\} \rangle_{0} = \delta_{ij}\delta_{\mathbf{k}\mathbf{p}}\delta_{\sigma\sigma'} G_{i}(\mathbf{k},\tau-\tau'),
	\label{internal_GF_definition}
\end{equation}
where $G_{i}(\mathbf{k},\tau-\tau')$ is the electron Green's function on grain $i$. We also noted that the off-diagonal terms, $\alpha \neq \beta$, are zero since Wick's theorem requires hopping between the same two grains $i$ and $i\pm\alpha$ in the two electron Green's functions.

\begin{figure}[t]
	\centering
	\includegraphics[width=0.3\linewidth]{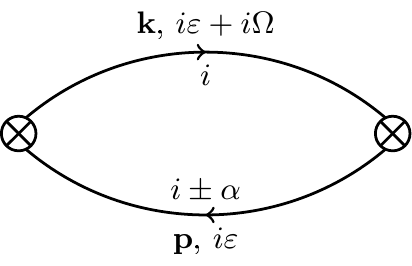}
	\caption{Diagrammatic representation of the Drude linear response function in eq. \ref{linear_response_function}.
                     The solid lines are the impurity and tunnelling averaged electron Green's functions; the crossed circles represent the tunnelling
                     matrix elements which arise from the tunnelling current.}
	\label{Granular_Drude_diagram}
\end{figure}

We represent $\mathcal{G}_{\alpha\alpha}$ diagrammatically in fig. \ref{Granular_Drude_diagram}, where the crossed circles are the tunnelling events associated to the current vertices, which carry a factor of $aet_{ij}^{\mathbf{k}\mathbf{p}}/\sqrt{\mathcal{N}}$, and the Green's function lines have a grain label in addition to the standard (intragrain) momentum and Matsubara frequency labels. The result of averaging over tunnelling events is to correlate the 
two current vertices, such that the correlated matrix elements produce a factor of $t^{2}a^{d}$. This factor is analogous to the $(2\pi N(0)\tau_{0})^{-1}$
factor created by correlated impurity scattering events in the homogeneous case.
	
\begin{figure}[b]
	\centering
	\includegraphics[width=0.95\linewidth]{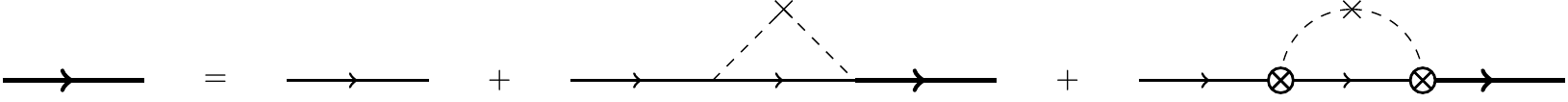}
	\caption{The diagrammatic series for the electron Green's function in a granular system. The additional term relative to fig. \ref{impurity_GF_series}
                     is due to correlated tunnelling back and forth between neighbouring grains.}
	\label{electron_GF_diagram}
\end{figure}
	
The electron Green's function can be represented by the diagrammatic series shown in fig. \ref{electron_GF_diagram}, and has the form
\begin{equation}
	G_{i}(\mathbf{k},i\varepsilon) = \frac{1}{i\varepsilon - \xi_{\mathbf{k}} + \frac{i}{2\tau}\text{sgn}(\varepsilon)},
	\label{electron_GF}
\end{equation}
where $\tau^{-1} = \tau_{0}^{-1} + z\Gamma$. $\tau_{0}^{-1}$ is the intragrain scattering rate due to impurities, and $z=2d$ is the coordination 
number of the grains. The tunnelling rate across a single junction, $\Gamma$, is related to the tunnelling amplitude via Fermi's golden rule,
\begin{equation}
	\Gamma = 2\pi N(0)t^{2}a^{d}.
	\label{Fermis_golden_rule}
\end{equation}
	
Unsurprisingly issues of convergence arise again, and we cannot freely interchange summation over momentum and frequency -- indeed the situation
is worse than for the homogeneous system as we now have two momentum sums. As before we will proceed by initially naively performing the momentum
sums first, before carrying out a more rigorous treatment of the swapping of summation order. We perform these calculations in granular real space as opposed to lattice momentum space; the latter formulation of this problem is given in appendix B.
	
\subsection{The Naive Approach} \label{Naive_granular}
	
In this section we arbitrarily swap the orders of momentum and frequency summations, and approximate the momentum sums by energy integrals 
around the Fermi surface,
\begin{equation}
	\mathcal{G}_{\alpha\alpha}(i\Omega) = 4e^{2}t^{2}N(0)^{2}a^{d+2} T\sum_{\varepsilon} 
        \int_{-\infty}^{+\infty} \frac{d\xi}{i\varepsilon - \xi + \frac{i}{2\tau}\text{sgn}(\varepsilon)}
        \int_{-\infty}^{+\infty} \frac{d\xi'}{i\varepsilon + i\Omega - \xi' + \frac{i}{2\tau}\text{sgn}(\varepsilon+\Omega)}.
	\label{granular_swapped_orders}
\end{equation}
The energy integrals are straightforward to perform and yield
\begin{equation}
	\mathcal{G}_{\alpha\alpha}(i\Omega) = 4e^{2}t^{2}N(0)^{2}a^{d+2} T\sum_{\varepsilon} 
                                                                   \left[-\pi^{2}\text{sgn}(\varepsilon)\text{sgn}(\varepsilon+\Omega)\right].
	\label{granular_non_kosher_post_momentum}
\end{equation}
To perform the frequency sum we note that $\text{sgn}(\varepsilon)\text{sgn}(\varepsilon+\Omega)=1-2\Theta[-\varepsilon(\varepsilon+\Omega)]$,
and that the sum over the Heaviside function term gives
\begin{equation}
	\mathcal{G}_{\alpha\alpha}(i\Omega) = 4e^{2}t^{2}N(0)^{2}a^{d+2}\left[2\pi^2 T \frac{\Omega}{2\pi T}\right]
                                                                = 4\pi e^{2} t^{2}N(0)^{2}a^{d+2}\Omega,
        \label{granular_non_kosher_partial_frequency}
\end{equation}
which clearly reproduces the granular Drude conductivity $\sigma_0^T=2e^{2}N(0)\Gamma a^2$. However, we still have to consider the 
apparently divergent sum of unity over all $\varepsilon$, and provide an argument that it should vanish. To do this, consider the more general sum,
\begin{equation}
	T\sum_{\varepsilon} \frac{1}{|\varepsilon|^m}=\frac{T}{(2\pi T)^m}\sum_{n=-\infty}^{+\infty} \frac{1}{|n+\alpha|^{m}}
                                                                               \equiv\frac{T}{(2\pi T)^m}S(m,\alpha),
        \label{vanishing_sum_definition}
\end{equation}
where $\alpha=\frac{1}{2}$ here. The sum $S(m,\alpha)$ can be written as
\begin{equation}
        S(m,\alpha)=\zeta(m,\alpha)+\zeta(m,-\alpha)-\frac{1}{\alpha^m},
        \label{vanishing_sum_zeta_functions}
\end{equation}
where the Hurwitz zeta function \cite{Gradshteyn_Ryzhik} is defined by
\begin{equation}
        \zeta(m,\alpha)=\sum_{n=0}^{\infty} \frac{1}{(n+\alpha)^m}.
        \label{Hurwitz_zeta_function}
\end{equation}
If we now take the limit $m\rightarrow 0$, and use the result \cite{Gradshteyn_Ryzhik}
\begin{equation}
	\zeta(0,\alpha) = \textstyle\frac{1}{2} - \alpha,
	\label{zeta_zero_definition}
\end{equation}
we see that
\begin{equation}
	S(0,\alpha) = \left(\textstyle\frac{1}{2} - \alpha\right) + \left(\textstyle\frac{1}{2} + \alpha\right) - 1 = 0.
	\label{vanishing_sum_result}
\end{equation}
An argument can therefore be made on the grounds of analytic continuation that the sum of unity over all fermionic Matsubara frequencies 
$\varepsilon$ vanishes, although this is not particularly rigorous. The reason for keeping $\alpha$ general is that the above analysis shows 
that the sum of unity over all bosonic Matsubara frequencies also vanishes.

As a final note, we see that the Drude conductivity has no $\Omega$ dependence, unlike the case of the homogeneous metal. It is possible
that this is an artefact of the naive approach taken in this section, so we should check whether this lack of frequency dependence remains in
the more rigorous approach which follows.
	
\subsection{A Careful Treatment}
We convert the Matsubara frequency sum in eq. \ref{linear_response_function} in exactly the same way as for eq. \ref{Drude_diagram_GFs_full},
and analytically continue $i\Omega\rightarrow\omega+i\delta$ to obtain
\begin{equation}
\begin{split}
        \mathcal{G}_{\alpha\alpha}^{R}(\omega) = &\frac{e^{2}N(0)i\Gamma a^{2}}{\pi^{2}} \int_{-\infty}^{+\infty} d\xi \int_{-\infty}^{+\infty} d\xi' \\
	&\times \int_{-\infty}^{+\infty} dz f(z) \left[G^{R}(\xi',z) - G^{A}(\xi',z)\right]\left[G^{R}(\xi,z+\omega) + G^{A}(\xi,z-\omega)\right].
	\label{analytic_continuation}
\end{split}
\end{equation}
We first demonstrate that $\mathcal{G}_{\alpha\alpha}(0)=0$ by using standard integration by parts methods \cite{Rickayzen}, starting with the triple
integral in the form
\begin{equation}
       -\frac{2i}{\tau} \int_{-\infty}^{+\infty} d\xi \int_{-\infty}^{+\infty} d\xi' \int_{-\infty}^{+\infty} dz f(z) (z-\xi) g(z-\xi) g(z-\xi'),
\end{equation}
where
\begin{equation}
	g(z-\xi) = \frac{1}{(z-\xi)^{2} + \frac{1}{4\tau^{2}}}.
	\label{GF_combination_denominator}
\end{equation}
Integrating by parts over $\xi'$, noting that the surface term vanishes, and that the derivative with respect to $\xi'$ may be replaced by a derivative
with respect to $-z$, we obtain 
\begin{equation}
	 -\frac{2i}{\tau} \int_{-\infty}^{+\infty} d\xi \int_{-\infty}^{+\infty} d\xi' \xi' \int_{-\infty}^{+\infty} dz f(z) (z-\xi) g(z-\xi) \frac{d}{dz}g(z-\xi').
	\label{leading_order_IBP_first}
\end{equation}
We now integrate by parts with respect to $z$, again noting that the surface term vanishes, to get
\begin{equation}
	\frac{2i}{\tau}\int_{-\infty}^{+\infty} d\xi \int_{-\infty}^{+\infty}  d\xi' \xi' \int_{-\infty}^{+\infty}  dz \, g(z-\xi') 
        \bigg[\frac{df}{dz}(z-\xi)g(z-\xi) + f(z)\frac{d}{dz}\{(z-\xi)g(z-\xi)\} \bigg].
	\label{leading_order_IBP_second}
\end{equation}
We may freely swap the orders of integration in the first term of eq. \ref{leading_order_IBP_second}, allowing us to perform the $\xi$ integral first; 
this term is then trivially zero as the integrand is odd in $z-\xi$. The second term can be computed by replacing the $z$ derivative by a $-\xi$ derivative, 
which can then be taken outside the $\xi'$ and $z$ integrals. Performing the $\xi$ integral then yields a boundary term which vanishes. It follows
that $\mathcal{G}_{\alpha\alpha}(0) = 0$, as we would expect for a system in the normal state.
	
In order to evaluate $\sigma_{0}^{T} = \sigma^{T}(\omega = 0)$, we expand $G_{\alpha\alpha}(\omega)$ to $\mathcal{O}(\omega)$, and make use of the 
relation $\mathcal{G}_{\alpha\alpha}(\omega) = -i\omega\sigma^{T}(\omega)$, to obtain
\begin{equation}
        \sigma_{0}^{T} = \frac{e^{2}N(0)\Gamma a^{2}}{\pi^{2}\tau^{2}} \int_{-\infty}^{+\infty} d\xi \int_{-\infty}^{+\infty} 
                                     d\xi' \int_{-\infty}^{+\infty} dz f(z) g(z-\xi')\frac{d}{dz}g(z-\xi).
	\label{Drude_integrals}
\end{equation}
We can write this as one half of a symmetric sum which includes the term with $\xi\leftrightarrow\xi'$, and then use integration by parts to move the 
$z$-derivative onto $f(z)$, giving
\begin{equation}
	\sigma_{0}^{T} = -\frac{e^{2}N(0)\Gamma a^{2}}{2\pi^{2}\tau^{2}} \int_{-\infty}^{+\infty} d\xi \int_{-\infty}^{+\infty} 
                                    d\xi' \int_{-\infty}^{+\infty} dz \frac{df}{dz} g(z-\xi') g(z-\xi).
	\label{Drude_IBP}
\end{equation}
The integrand now falls off sufficiently quickly to allow the orders of integration to be swapped, allowing us to perform the $\xi$ and $\xi'$ integrals first. 
The $\xi$ and $\xi'$ integrals both give a factor $2\pi\tau$, whilst the $z$-integral gives factor $-1$, leading to the final result for the granular Drude
conductivity
\begin{equation}
	\sigma_{0}^{T} = 2e^{2}N(0)\Gamma a^{2}.
	\label{granular_Drude_conductivity_correct}
\end{equation}
	
To find the finite-frequency response of the system, we expand eq. \ref{linear_response_function} in powers of $\omega$. The coefficient
of $\omega^n$ in the expansion of $\sigma^T(\omega)$ is then
\begin{equation}
\begin{split}
	\sigma_{n}^{T} = \frac{e^{2}N(0)\Gamma a^{2}}{\pi^{2}\tau} 
                                    \int_{-\infty}^{+\infty} d\xi' \int_{-\infty}^{+\infty} &d\xi \int_{-\infty}^{+\infty} dz 
                                    \frac{f(z)}{(\xi'-z)^{2}+\frac{1}{4\tau^{2}}}\\[5pt]
                                    &\times \left[\frac{(-1)^{n+1}}{(z-\xi+\frac{i}{2\tau})^{n+2}}+\frac{1}{(z-\xi-\frac{i}{2\tau})^{n+2}}\right].
	\label{AC_first_order_start}
\end{split}
\end{equation}
We can integrate by parts with respect to $\xi$, noting that the boundary term vanishes, and replace the $\xi$ derivative by a $-z$ derivative.
Integrating by parts with respect to $z$ then gives
\begin{equation}
\begin{split}
	\sigma_{n}^{T} = -\frac{e^{2}N(0)\Gamma a^{2}}{\pi^{2}\tau} 
                                    \int_{-\infty}^{+\infty} d\xi' &\int_{-\infty}^{+\infty} d\xi \xi \int_{-\infty}^{+\infty} dz 
                                    \left[\frac{(-1)^{n+1}}{(z-\xi+\frac{i}{2\tau})^{n+2}}+\frac{1}{(z-\xi-\frac{i}{2\tau})^{n+2}}\right]\\[5pt]
                                    &\times\left[\frac{1}{(\xi'-z)^{2}+\frac{1}{4\tau^{2}}}{df\over dz}
                                    +f(z){d\over dz}\left(\frac{1}{(\xi'-z)^{2}+\frac{1}{4\tau^{2}}}\right)\right].
	\label{AC_first_order_by_parts}
\end{split}
\end{equation}
In the first term, the $df/dz$ factor allows us to swap the orders of integration; the $\xi$ integral then yields zero as we can close the contour
for each term so that no poles are enclosed. In the second term we may replace the $z$ derivative by a $-\xi'$ derivative, and then take this
derivative outside the $z$ and $\xi$ integrals. The $\xi'$ integral then only gives a contribution from the boundaries which is zero. It follows
that $\sigma_{n}^{T}=0$ for all $n\ge 1$, and hence the tunnelling conductivity has no frequency dependence, reproducing the result found in 
section \ref{Naive_granular}.

As in the naive approach, we see that the Drude conductivity for granular metals is independent of $\Omega$. This difference from the
homogeneous result is due to the momentum scrambling which takes place in the tunnelling between grains, which leads to the two
independent momentum integrals. The frequency-dependent conductivity in granular metals will come from interference phenomena
such as weak localisation \cite{Biagini2005_WL} and electron-electron interaction \cite{Efetov2003} effects.

\section{Conclusions} \label{Conclusions_sec}
	
We have provided a detailed analysis of the diagrammatic method for calculating the electrical conductivity in disordered granular metals. 
In doing so, we find that the granular Drude conductivity suffers from more severe convergence issues around swapping orders of frequency
and momentum integrals compared to the homogeneous metal. These issues can be resolved using the same integration by parts techniques as
in the homogeneous case, although the presence of an extra momentum integral makes these slightly more involved. We find that naively swapping
the order of summation gives the correct result if we interpret a divergent frequency sum in an appropriate manner using analytic continuation via the
Hurwitz zeta function. In contrast to the homogeneous case, there is no finite diamagnetic term which needs to be cancelled by a term arising from
the current-current correlator. We demonstrate that the electromagnetic response function vanishes in the zero frequency limit, as expected for
a normal (non-superconducting) system, and derive the correct expression for the granular Drude conductivity expected from the Einstein relation
and Fermi's golden rule. To the best of our knowledge, this is the first explicit diagrammatic calculation of these results.

The fact that naively swapping the order of summation gives the correct result is beneficial for calculations, as it is mathematically simpler to perform 
momentum integrals first, followed by Matsubara frequency sums, which avoids the necessity of analytic continuation in the complex frequency
plane. We note also that no frequency dependence occurs in the granular Drude conductivity, unlike in the homogeneous metal. This is due to the
intragrain momentum scrambling which occurs in the tunnelling events. This is not physically relevant since the dominant frequency dependence
will come from interference phenomena such as weak localisation \cite{Biagini2005_WL} and electron-electron interaction \cite{Efetov2003} effects.
	
\appendix
\gdef\thesection{Appendix \Alph{section}:}
	
\section{Tight Binding Current Density Operator} 
\label{appendix_current_operator}

In this appendix we provide a derivation of the granular current density operator, starting from the discretised version of the continuity equation.
The latter can be written as
	\begin{equation}
		\frac{d\rho_{k}}{dt} = -\nabla \cdot \mathbf{j}_{k},
		\label{continuity_equation}
	\end{equation}
where $\rho_{k} = ec^{\dagger}_{k}c_{k}/a^{d}$ and $j_{k}$ are the charge density and current density on site $k$, respectively. 
To find the time derivative of $\rho_{k}$ we use the Heisenberg equation of motion,
	\begin{equation}
		\frac{d\rho_{k}}{dt} = \frac{ie}{a^{d}} \big[H_{T},c_{k}^{\dagger}c^{\null}_{k}\big],
		\label{Heisenberg_EOM}
	\end{equation}
where $H_{T}$ is the tight binding Hamiltonian in the presence of a vector potential $\mathbf{A}$. This is given by the second term in eq. \ref{full_Hamiltonian}. Since
	\begin{equation}
		\big[c_{i}^{\dagger}c^{\null}_{j},c_{k}^{\dagger}c^{\null}_{k}\big] 
                = \delta_{jk}c_{i}^{\dagger}c^{\null}_{k} - \delta_{ik}c_{k}^{\dagger}c^{\null}_{j},
		\label{charge_density_commutator}
	\end{equation}
it follows that
	\begin{equation}
	\begin{split}
		\frac{d\rho_{k}}{dt} &= \frac{ie}{a^{d}} \sum_{i,j} t_{ij} e^{ie\mathbf{A}\cdot\mathbf{R}_{ij}} 
                \left(\delta_{jk}c_{i}^{\dagger}c^{\null}_{k} - \delta_{ik}c_{k}^{\dagger}c^{\null}_{j}\right) \\
		&= \frac{ie}{a^{d}} \sum_{\alpha} \left(t_{k+\alpha,k}e^{ieaA_{\alpha}}c_{k+\alpha}^{\dagger}c^{\null}_{k} 
                   + t_{k-\alpha,k}e^{-ieaA_{\alpha}}c_{k-\alpha}^{\dagger}c^{\null}_{k} - \hbox{H.c.}\right)
		\label{charge_density_derivative}
	\end{split}
	\end{equation}
where in the second equality we made use of eq. \ref{tunnelling_matrix_elements} and $t_{ij} = t_{ji}^{*}$. The sum over $\alpha$ is the sum over
the $d$ directions in the lattice. 
	
	\begin{figure}[t]
		\centering
		\includegraphics[width=0.3\linewidth]{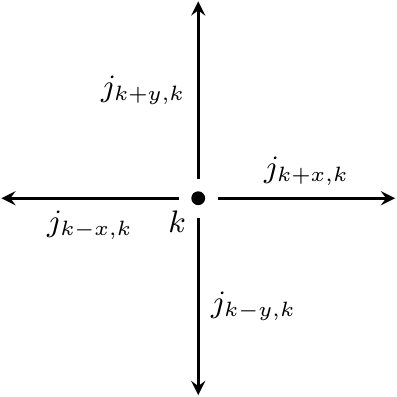}
		\caption{Discretised divergence of the vector $\mathbf{j}$ at the point $k$ on a 2D square lattice.}
		\label{discrete_divergence_image}
	\end{figure}
	
We now address the issue of discretising the divergence of the current operator. The divergence may be thought of as the sum of the current moving 
away from the lattice point $k$, as shown in fig. \ref{discrete_divergence_image} for  a 2D square lattice. We can therefore infer that
	\begin{equation}
		\nabla \cdot \mathbf{j}_{k} = \frac{1}{a} \sum_{\alpha} \left( j_{k+\alpha,k}+j_{k-\alpha,k} \right),
		\label{discrete_divergence}
	\end{equation}
where $j_{k\pm\alpha,k}$ is the current moving from lattice site $k$ to $k\pm\alpha$. By combining eq. \ref{charge_density_derivative} and 
eq. \ref{discrete_divergence} with the continuity equation, we obtain the tight binding current density operator component from site $k$ to $k+\alpha$,
	\begin{equation}
		j_{k+\alpha,k} = -\frac{ie}{a^{d-1}}\left( t_{k+\alpha,k}e^{ieaA_{\alpha}}c_{k+\alpha}^{\dagger}c^{\null}_{k} 
                                          - t_{k,k+\alpha}e^{-ietaA_{\alpha}}c_{k}^{\dagger}c^{\null}_{k+\alpha} \right).
		\label{tight_binding_current_density_operator}
	\end{equation}
To obtain eq. \ref{full_current_operator} for the macroscopic current density $j_\alpha$, we average this expression over all sites $k$.

\section{Lattice Momentum Space} \label{appendix_lattice_momentum}
In the following we perform the granular Drude calculation in lattice momentum space, which is physically sensible as the system possesses lattice 
translational invariance after averaging over the distribution of tunnelling matrix elements. Most works using diagrammatic methods for granular metals start
with the current density operator written in lattice momentum space \cite{Beloborodov2007,Biagini2005_WL}. To first order in $\mathbf{A}$ this operator is
\begin{equation}
	j_{\alpha} = -\frac{2et}{a^{d-1}\mathcal{N}} \sum_{\mathbf{K}} \sum_{\substack{\mathbf{k},\mathbf{p}, \\ \sigma}} \sin(K_{\alpha}a) c_{\mathbf{K}\sigma\mathbf{k}}^{\dagger} c_{\mathbf{K}\sigma\mathbf{p}}^{\null} + \frac{e^{2}A_{\alpha}}{a^{d-2}\mathcal{{N}}}H_{T}^{(0)},
\label{current_operator_lattice_momentum}
\end{equation}
where $\mathbf{K}$ is the lattice momentum. In general we will use lower case letters for intragranular momenta, and capital letters for intergranular momenta. In writing this operator we have disobeyed the assumption that the $t_{ij}^{\mathbf{k}\mathbf{p}}$ come from a random white-noise
distribution, and replace them by a non-random value $t$. We will discuss this discrepancy shortly.

Substituting the above into the current-current correlator gives,
\begin{equation}
	\mathcal{G}_{\alpha\beta}(i\Omega) = \frac{8e^{2}t^{2}}{a^{d-2} \mathcal{N}} \sum_{\mathbf{K}}\sum_{\mathbf{k},\mathbf{p}} 
        T\sum_{\varepsilon} \sin(K_{\alpha}a)\sin(K_{\beta}a) G(\mathbf{K},\mathbf{k},i\varepsilon) G(\mathbf{K},\mathbf{p},i\varepsilon+i\Omega),
	\label{current_current_correlator_lattice_momentum}
\end{equation}
where $G(\mathbf{K},\mathbf{k},i\varepsilon)$ is the electron Green's function describing motion within the grain and between the grains. 
We associate a factor of $2aet\sin(K_{\alpha}a)/\sqrt{\mathcal{N}}$ to the current vertices in lattice momentum space, and multiply the resulting sums by 
a factor of $a^{d}$ due to matrix element averaging. In writing eq. \ref{current_current_correlator_lattice_momentum}, we noted that the first two terms of
 eq. \ref{macroscopic_current} vanish in this picture. The $j_{0,\alpha}$ term carries a single factor of $\sin(K_{\alpha}a)$ which clearly goes to zero
 under a $\mathbf{K}$ summation. Similarly, the $H_{T}^{(0)}$ term carries a single factor of $\cos(K_{\alpha}a)$, which also vanishes under 
summation over $\mathbf{K}$.

Assuming that the intergranular momentum dependence of the electron Green's function may be ignored, we may compute the current-current correlator as,
\begin{equation}
	\mathcal{G}_{\alpha\beta}(i\Omega) = \delta_{\alpha\beta} \frac{4e^{2}t^{2}}{a^{d-2}} \sum_{\mathbf{k},\mathbf{p}} 
        T\sum_{\varepsilon} G(\mathbf{k},i\varepsilon) G(\mathbf{p},i\varepsilon+i\Omega),
	\label{current_current_correlator_lattice_momentum_simplified}
\end{equation}
where we noted that $\sum_{\mathbf{K}} \sin(K_{\alpha}a)\sin(K_{\beta}a) = \delta_{\alpha\beta} \mathcal{N}/2$. The subsequent analysis is then 
identical to the lattice real space picture discussed previously.

The only way we can justify replacing the random matrix elements $t_{ij}^{\mathbf{k}\mathbf{p}}$ with a constant non-random value $t$, is if we
can prove that this leads to the same result after averaging over the random distribution. To investigate this question, we consider the application
of Wick's theorem to the current-current correlator in eq. \ref{Matsubara_currect_current_correlator}. If we make the assumption
$\langle t_{ij}^{\mathbf{k}\mathbf{p}} \rangle = t \delta_{j,i\pm\alpha}$ for the current vertex alone, we obtain the expression
\begin{equation}
\begin{split}
	\sum_{i,j}\sum_{\sigma,\sigma'}\sum_{\mathbf{k},\mathbf{p}}\sum_{\mathbf{k}',\mathbf{p}'} \Big[ &t^{2}\langle 
         T_{\tau}\{c_{i+\alpha\sigma\mathbf{k}}^{\dagger}(\tau)c_{i\sigma\mathbf{p}}^{\null}(\tau)c_{j+\beta\sigma'\mathbf{k}'}^{\dagger}(\tau')c_{j\sigma'\mathbf{p}'}^{\null}(\tau')\} \rangle_{0} \\[-5pt]
	&\qquad + t^{2}\langle T_{\tau}\{c_{i\sigma\mathbf{k}}^{\dagger}(\tau)c_{i+\alpha\sigma\mathbf{p}}^{\null}(\tau)c_{j\sigma'\mathbf{k}'}^{\dagger}(\tau')c_{j+\beta\sigma'\mathbf{p}'}^{\null}(\tau')\} \rangle_{0} \\[3pt]
	&\qquad\qquad -t^{2}\langle T_{\tau}\{c_{i+\alpha\sigma\mathbf{k}}^{\dagger}(\tau)c_{i\sigma\mathbf{p}}^{\null}(\tau)c_{j\sigma'\mathbf{k}'}^{\dagger}(\tau')c_{j+\beta\sigma'\mathbf{p}'}^{\null}(\tau')\} \rangle_{0} \\[3pt]
	&\qquad\qquad\qquad - t^{2}\langle T_{\tau}\{c_{i\sigma\mathbf{k}}^{\dagger}(\tau)c_{i+\alpha\sigma\mathbf{p}}^{\null}(\tau)c_{j+\beta\sigma'\mathbf{k}'}^{\dagger}(\tau')c_{j\sigma'\mathbf{p}'}^{\null}(\tau')\} \rangle_{0} \Big].
\label{current_current_pre_Wicks_theorem}
\end{split}
\end{equation}
Transforming to lattice momentum space, this becomes
\begin{equation}
\begin{split}
	t^{2}\sum_{\mathbf{K},\mathbf{Q}}\sum_{\sigma,\sigma'}\sum_{\mathbf{k},\mathbf{p}}\sum_{\mathbf{k}',\mathbf{p}'} \Big[\Big( &e^{-ia(K_{\alpha}+Q_{\beta})} + e^{ia(K_{\alpha}+Q_{\beta})} - e^{-ia(K_{\alpha}-Q_{\beta})} - e^{ia(K_{\alpha}-Q_{\beta})}\Big) \\[-8pt]
	&\qquad\times \langle T_{\tau}\{c_{\mathbf{K}\sigma\mathbf{k}}^{\dagger}(\tau)c_{\mathbf{K}\sigma\mathbf{p}}^{\null}(\tau)c_{\mathbf{Q}\sigma'\mathbf{k}'}^{\dagger}(\tau')c_{\mathbf{Q}\sigma'\mathbf{p}'}^{\null}(\tau')\} \rangle_{0}\Big] \\[8pt]
	= -4t^{2}\sum_{\mathbf{K},\mathbf{Q}}\sum_{\sigma,\sigma'}\sum_{\mathbf{k},\mathbf{p}}\sum_{\mathbf{k}',\mathbf{p}'} &\sin(K_{\alpha}a)\sin(Q_{\beta}a) \langle T_{\tau}\{c_{\mathbf{K}\sigma\mathbf{k}}^{\dagger}(\tau)c_{\mathbf{K}\sigma\mathbf{p}}^{\null}(\tau)c_{\mathbf{Q}\sigma'\mathbf{k}'}^{\dagger}(\tau')c_{\mathbf{Q}\sigma'\mathbf{p}'}^{\null}(\tau')\} \rangle_{0}
\label{current_current_pre_Wicks_lattice_momentum}
\end{split}
\end{equation}
To obtain eq. \ref{current_current_correlator_lattice_momentum} from eq. \ref{current_current_pre_Wicks_lattice_momentum},
we apply Wick's theorem in conjunction with
\begin{equation}
	-\langle T_{\tau}\{c_{\mathbf{K}\sigma\mathbf{k}}^{\dagger}(\tau)c_{\mathbf{Q}\sigma'\mathbf{p}}^{\null}(\tau')\} \rangle_{0} = \delta_{\mathbf{K}\mathbf{Q}}\delta_{\sigma\sigma'}\delta_{\mathbf{k}\mathbf{p}} G(\mathbf{K},\mathbf{k},\tau-\tau'),
\label{lattice_momentum_space_GF_correlator}
\end{equation}
and finally note that $G(\mathbf{K},\mathbf{k},i\varepsilon)=G(\mathbf{k},i\varepsilon)$ does not depend upon $\mathbf{K}$. We therefore
see that we obtain the same result if we replace the random $t_{ij}^{\mathbf{k}\mathbf{p}}$ by constant non-random $t$ in the current vertex. 
In effect, since we will always get $t^2$ from the average of the two vertices, this amounts to assigning a non-random $t$ to each vertex. It is
not clear that this trick will always work for more complicated diagrams. For the purposes of careful calculation, it is better to work in granular
real space, and then only move to lattice momentum space after the average over the tunnelling matrix distribution has been taken.

\section*{Acknowledgements}
The authors are grateful to J M Fellows and G M Klemencic for useful discussions of transport phenomena in granular materials.

\end{document}